\documentclass[12pt]{article}
\usepackage{graphicx}
\usepackage{epsfig}
\usepackage{latexsym}
\usepackage{latexsym}
\usepackage{amssymb}
\usepackage{amsmath}

\input{tcilatex}

\begin{document}

\author{Khireddine Nouicer\thanks{%
Email: khnouicer@mail.univ-jijel.dz / khnouicer@yahoo.fr} \\
\textit{Laboratory of Theoretical Physics and Department of Physics,}\\
\textit{\ Faculty of Sciences, University of Jijel}\\
\textit{\ Bp 98, Ouled Aissa, 18000 Jijel, Algeria.}}
\title{Quantum-corrected black hole thermodynamics to all orders in the
Planck length }
\date{}
\maketitle

\begin{abstract}
We investigate the effects to all orders in the Planck length from a
generalized uncertainty principle (GUP) on black holes thermodynamics. We
calculate the corrected Hawking temperature, entropy, and examine in details
the Hawking evaporation process. As a result, the evaporation process is
accelerated and the evaporation end-point is a zero entropy, zero heat
capacity and finite non zero temperature black hole remnant (BHR). In
particular we obtain a drastic reduction of the decay time, in comparison
with the result obtained in the Hawking semi classical picture and with the
GUP to leading order in the Planck length.


\end{abstract}

\section{Introduction}

Recently a great interest has been devoted to the study of the effects of
generalized uncertainty principles (GUPs) and modified dispersion relations
(MDRs) on black holes thermodynamics. The concepts of GUPs and MDRs
originates from several studies in string theory approach to quantum gravity 
$\left[ \text{1- 4}\right] $, loop quantum gravity \cite{Garay},
noncommutative space-time algebra $\left[ \text{6 - 8}\right] $ and black
holes gedanken experiments $\left[ \text{9 - 10}\right] $. All these
approaches indicate that the standard Heisenberg uncertainty principle must
be generalized to incorporate additional uncertainties when quantum
gravitational effects are taken into account. Actually it is believed that
any promising candidate for a quantum theory of gravity \ must include the
GUPs and/or MDRs \ as central ingredients.

The main consequence of the GUP is the appearance of a minimal length scale
of the order of the Planck length which cannot be probed, providing a
natural UV cut-off, and thus corrections to black holes thermodynamic
parameters\ are expected at the Planck scale.

The consequences of GUPs and/or MDRs on black holes thermodynamics have been
considered intensively in the recent literature on the subject $\left[ \text{%
11 - 16}\right] $. Notably, it has been shown that GUP prevents black holes
from complete evaporation, exactly like the standard Heisenberg principle
prevents the hydrogen atom from total collapse \cite{adler}. Then at the
final stage of the Hawking radiation process of a black hole, a inert black
hole remnant (BHR) continue to exist with zero entropy, zero heat capacity
and a finite non zero temperature. The inert character of the BHR, besides
gravitational interactions, makes this object a serious candidate to explain
the nature of dark matter \cite{Pisin,P-adler}. On the other hand, a
particular attention has been also devoted to the computation of the entropy
of a black hole and the sub-leading logarithmic correction $\left[ \text{20
- 34}\right] $.

All the above studies have been performed with a GUP to leading order in the
Planck length. However, recent generalization of the GUP induces
quantitative corrections to the entropy and then influences the evaporation
phase of the black hole \cite{ko}. Besides this growing interest in quantum
gravity phenomenology, a intense activity is actually devoted to possible
production of black holes at particle colliders \cite{cern1,cern2} and in
ultrahigh energy cosmic ray (UHECR) airshowers \cite{feng,c}. The next
generation of particle colliders are planned to reach a \textit{c-m} energy
of the order of few \textit{TeV} , a scale at which the complete evaporation
of BH is expected to end, leaving up in a scenario with GUP a inert BHR.
Then, it is phenomenologically relevant, to obtain the corrections to BH
thermodynamic parameters in the framework of a GUP beyond the leading order
in the Planck length.

In this paper we discuss the effects, brought by a generalization of the GUP
to all orders in the Planck length, on thermodynamic parameters of the
Schwarzschild black hole . Hereafter, we refer to this version of GUP as GUP$%
^{\ast }$.

The organization of this work is as follows. In section 2, we introduce a
deformed position and momentum operators algebra leading to GUP$^{\ast }$
and examine its various implications. In section 3, the Hawking temperature
and entropy are computed and the departures from the standard case shown. In
section 4, we calculate the deviation from the standard Stefan-Boltzmann law
of the black body radiation spectrum and investigate the Hawking evaporation
process of black holes by a calculation of the evaporation rate, the decay
time and the heat capacity. Finally we compare our results with the ones
obtained in the context of the GUP to leading order in the Planck length
commonly used in the literature. Our conclusions are summarized in the last
section.

\section{Generalized uncertainty principle}

Loop quantum gravity and string theory approach to quantum gravity predict
slight deviations in the laws describing photons propagation in vacuum. It
is expected that these effects, leading to a modified dispersion relation
(MDR), could be amplified by cosmological distances and then become
observables \cite{ahluwalia}. On the other hand,  quantum gravity
phenomenology has been tackled within effective models based on MDRs and/or
GUPs and containing the minimal length as a natural UV cut-off. Recently the
relation between these approaches has been clarified and established \cite%
{Sabine}.

The idea of a minimal length can be modelled in terms of a quantized
space-time and goes back to the early days of quantum field theory \cite%
{snyder} (see also $\left[ 40-43\right] $ ). An other approach is to
consider deformations to the standard Heisenberg algebra \cite{c222,mang},
which lead to generalized uncertainty principles. In this section we follow
the latter approach and exploit results recently obtained. Indeed, it has
been shown in the context of canonical noncommutative field theory in the
coherent states representation \cite{spallucci} and field theory on
non-anticommutative superspace \cite{Moffat,nouicer}, that the Feynman
propagator display an exponential UV cut-off of the form $\exp \left( -\eta
p^{2}\right) $, where the parameter $\eta $ is related to the minimal
length. This framework has been further applied, in series of papers \cite%
{nicolini}, to the black hole evaporation process.

At the quantum mechanical level, the essence of the UV finiteness of the
Feynman propagator can be also captured by a non linear relation, $p=f(k)$,
between the momentum and the wave vector of the particle \cite{Sabine}. This
relation must be invertible and has to fulfill the following requirements:

\begin{enumerate}
\item For smaller energies than the cut-off the usual dispersion relation is
recovered.

\item For large energies, the wave vector asymptotically reaches the cut-off.
\end{enumerate}

In this case, the usual momentum measure $d^{n}p$ is deformed and becomes $%
d^{n}p\prod_{i}\frac{\partial k_{i}}{\partial p_{j}}$. In the following, we
will restrict ourselves to the isotropic case and work with one space-like
dimension. Following \cite{spallucci,nouicer} and setting $\eta =\frac{%
\alpha ^{2}L_{Pl}^{2}}{\hbar ^{2}}$ we have 
\begin{equation}
\frac{\partial p}{\partial k}=\hbar \text{exp}\left( \frac{\alpha
^{2}L_{Pl}^{2}}{\hbar ^{2}}p^{2}\right) ,  \label{measure}
\end{equation}%
where $\alpha $ is a dimensionless constant of order one.

From Eq.$\left( \ref{measure}\right) $ we obtain the dispersion relation 
\begin{equation}
k\left( p\right) =\frac{\sqrt{\pi }}{2\alpha L_{Pl}}\text{erf}\left( \frac{%
\alpha L_{Pl}}{\hbar }p\right) ,
\end{equation}%
from which we have the following minimum Compton wavelength 
\begin{equation}
\lambda _{0}=4\sqrt{\pi }\alpha L_{Pl}.  \label{compton}
\end{equation}

Let us show that these results can be obtained from the following
representation of the position and momentum operators \ 
\begin{equation}
X=i\hbar \exp \left( \frac{\alpha ^{2}L_{Pl}^{2}}{\hbar ^{2}}P^{2}\right) {%
\partial _{p}}\qquad P=p.  \label{xp}
\end{equation}

The corrections to the standard Heisenberg algebra become effective in the
so-called quantum regime where the momentum and length scales are of the
order of the Planck mass $M_{Pl}$ and of the Planck length $L_{Pl}$
respectively.

The hermiticity condition of the position operator implies the following
modified completeness relation  
\begin{equation}
\int dpe^{-\frac{\alpha ^{2}L_{Pl}^{2}}{\hbar ^{2}}p^{2}}|p\rangle \langle
p|=1  \label{ferm}
\end{equation}%
and  modified scalar product 
\begin{equation}
\left\langle p\right\vert \left. p^{\prime }\right\rangle =e^{\frac{\alpha
^{2}L_{\text{Pl}}^{2}}{\hbar ^{2}}p^{2}}\delta \left( p-p^{\prime }\right) .
\end{equation}%
From Eq.$\left( \ref{ferm}\right) $ we observe that we have reproduced the
Gaussian damping factor in the Feynman propagator \cite{spallucci,nouicer}.
The algebra defined by Eq. $\left( \ref{xp}\right) $ leads to the following
generalized commutator and generalized uncertainty principle (GUP$^{\ast }$) 
\begin{equation}
\left[ X,P\right] =i\hbar \exp \left( \frac{\alpha ^{2}L_{Pl}^{2}}{\hbar ^{2}%
}P^{2}\right) ,\quad \left( \delta X\right) \left( \delta P\right) \geq 
\frac{\hbar }{2}\left\langle \exp \left( \frac{\alpha ^{2}L_{Pl}^{2}}{\hbar
^{2}}P^{2}\right) \right\rangle .  \label{GUP}
\end{equation}%
In order to investigate the quantum implications of this deformed algebra,
we consider the saturate GUP$^{\ast }$ and solve for $\left( \delta P\right) 
$. Using the property $\left\langle P^{2n}\right\rangle \geq \left\langle
P^{2}\right\rangle ^{n}$ \ and $\left( \delta P\right) ^{2}=\left\langle
P^{2}\right\rangle -\left\langle P\right\rangle ^{2}$ the saturate GUP$%
^{\ast }$ is then given by

\begin{equation}
\left( \delta X\right) \left( \delta P\right) =\frac{\hbar }{2}\exp \left( 
\frac{\alpha ^{2}L_{P\text{l}}^{2}}{\hbar ^{2}}\left( \left( \delta P\right)
^{2}+\left\langle P\right\rangle ^{2}\right) \right) .
\end{equation}%
Taking the square of this expression we obtain

\begin{equation}
W\left( u\right) e^{W\left( u\right) }=u,  \label{lam}
\end{equation}%
where we have set $W(u)=-2\frac{\alpha ^{2}L_{Pl}^{2}}{\hbar ^{2}}\left(
\delta P\right) ^{2}$ and $u=-\frac{\alpha ^{2}L_{Pl}^{2}}{2\left( \delta
X\right) ^{2}}e^{-2\frac{\alpha ^{2}L_{P\text{l}}^{2}}{\hbar ^{2}}%
\left\langle P\right\rangle ^{2}}.$

The equation given by Eq.$\left(\ref{lam}\right)$ is exactly the definition
of the Lambert function \cite{Lambert}. The Lambert $W$ \ function is a
multivalued functions. Its different branches are labelled by the integer $%
k=0,\pm 1,\pm 2,\cdots $. When $u$ is a real number Eq.$\left(\ref{lam}%
\right)$ have two real solutions for $0\geq u\geq -\frac{1}{e}$, denoted by $%
W_{0}(u)$ and $W_{-1}(u)$, or it can have only one real solution for $u\geq
0 $, namely $W_{0}(u)$ . For -$\infty <u<-\frac{1}{e}$, Eq.(\ref{lam}) have
no real solutions.

Using Eq.(\ref{lam}) the uncertainty in momentum is then given by 
\begin{equation}
\left( \delta P\right) =\frac{\hbar e^{\frac{\alpha ^{2}L_{P\text{l}}^{2}}{%
\hbar ^{2}}\left\langle P\right\rangle ^{2}}}{2\left( \delta X\right) }\exp
\left( -\frac{1}{2}W\left( -\frac{\alpha ^{2}L_{Pl}^{2}e^{\frac{2\alpha
^{2}L_{Pl}^{2}}{\hbar ^{2}}\left\langle P\right\rangle ^{2}}}{2\left( \delta
X\right) ^{2}}\right) \right) .  \label{argu}
\end{equation}%
Then from the argument of the Lambert function in Eq.$\left( \ref{argu}%
\right) $ we have the following condition 
\begin{equation}
\frac{\alpha ^{2}L_{Pl}^{2}e^{\frac{2\alpha ^{2}L_{Pl}^{2}}{\hbar ^{2}}%
\left\langle P\right\rangle ^{2}}}{2\left( \delta X\right) ^{2}}\leqslant 
\frac{1}{e},
\end{equation}%
which leads to a minimal uncertainty in position given by 
\begin{equation}
\left( \delta X\right) _{\min }=\sqrt{\frac{e}{2}}\alpha L_{Pl}e^{\frac{%
\alpha ^{2}L_{Pl}^{2}}{\hbar ^{2}}\left\langle P\right\rangle ^{2}}.
\end{equation}%
The absolutely smallest uncertainty in position or minimal length \ is
obtained for physical states for which we have $\left\langle P\right\rangle
=0$ and $\left( \delta P\right) =\hbar /\left( \sqrt{2}\alpha L_{P\text{l}%
}\right) ,$ and is given by 
\begin{equation}
\left( \delta X\right) _{0}=\sqrt{\frac{e}{2}}\alpha L_{Pl}
\end{equation}%
In terms of the minimal length the momentum uncertainty becomes 
\begin{equation}
\left( \delta P\right) =\frac{\hbar }{2\left( \delta X\right) }\exp \left( -%
\frac{1}{2}W\left( -\frac{1}{e}\left( \frac{(\delta X)_{0}}{(\delta X)}%
\right) ^{2}\right) \right) .  \label{argup}
\end{equation}%
Here we observe that $\frac{1}{e}\frac{(\delta X)_{0}}{(\delta X)}<1$ is a
small parameter, by virtue of the GUP$^{\ast }$, and perturbative expansions
to all orders in the Planck length can be safely performed.

Indeed a series expansion of Eq.(\ref{argup}) gives the corrections to the
standard Heisenberg principle 
\begin{equation}
\delta P\simeq \frac{\hbar }{2\left( \delta X\right) }\left( 1+\frac{1}{2e}%
\left( \frac{(\delta X)_{0}}{(\delta X)}\right) ^{2}+\frac{5}{8e^{2}}\left( 
\frac{(\delta X)_{0}}{(\delta X)}\right) ^{4}+\frac{49}{48e^{3}}\left( \frac{%
(\delta X)_{0}}{(\delta X)}\right) ^{6}+\ldots \right) .
\end{equation}%
This expression of $\left( \delta P\right) $ containing only odd powers of $%
\left( \delta X\right) $ is consistent with a recent analysis in which
string theory and loop quantum gravity, considered as the most serious
candidates for a theory of quantum gravity, put severe constraints on the
possible forms of GUPs and MDRs \cite{c10}.

Before ending this section and for later use let us recall the form of the
GUP to leading order in the Planck length widely used in the literature on
quantum gravity phenomenology. This GUP is given by 
\begin{equation}
\left( \delta X\right) \left( \delta P\right) \geq \frac{\hbar }{2}\left( 1+%
\frac{\alpha ^{2}L_{Pl}^{2}}{\hbar ^{2}}\left( \delta P\right) ^{2}\right) .
\label{logup}
\end{equation}%
A simple calculation leads to the following minimal length 
\begin{equation}
\left( \delta X\right) _{0}=\alpha L_{Pl},
\end{equation}%
which is of order of the Planck length. However, as nicely noted in \cite%
{Sabine}, this form of GUP do not fulfill the second requirement listed
above. In the following sections we use the form of the GUP given by Eq.$%
\left( \ref{argup}\right) $ and investigate the thermodynamics of the
Schwarzschild black hole. We use units $\hbar =c=k_{\text{B}}=1$ which imply 
$L_{Pl}=M_{Pl}^{-1}=T_{Pl}^{-1}=\sqrt{G}.$

\section{Black hole thermodynamics}

The metric of a four-dimensional Schwarzschild black hole is given by

\begin{equation}
ds^{2}=\left( 1-\frac{2MG}{r}\right) dt^{2}-\left( 1-\frac{2M}{r}\right)
^{-1}dr^{2}-r^{2}d\Omega ^{2},
\end{equation}%
where $M$\ \ represents the mass of the black hole. The Schwarzschild
horizon radius, located at $r_{h}$, is defined by

\begin{equation}
r_{h}=2MG.  \label{r0}
\end{equation}%
Near-horizon geometry considerations suggests to set $\delta X\simeq r_{h}$,
and then Eq.$\left( \ref{r0}\right) $ leads to minimum horizon radius and
minimum mass given by

\begin{equation}  \label{min}
r_{h}=\left( \delta X\right) _{0}=\sqrt{\frac{e}{2}}\alpha L_{Pl},\quad
M_{0}=\frac{\alpha \sqrt{e}}{2\sqrt{2}}M_{Pl}.
\end{equation}%
Therefore, black holes with mass smaller than $M_{0}$ do not exist.

In the standard Hawking picture, temperature and entropy of the
Schwarzschild black hole of mass $M$ are \cite{a,b}

\begin{equation}
T_{\text{H}}=\frac{1}{8\pi GM},\quad S=4\pi GM^{2}.
\end{equation}%
Let us then examine the corrections to the above expressions due to the GUP$%
^{\ast }$. Following the heuristic argument of Bekenstein we have 
\begin{equation}
T_{\text{H}}\approx \frac{\delta P}{2\pi }.
\end{equation}%
Using Eq.$\left( \ref{argup}\right) $, the GUP$^{\ast }$-corrected Hawking
temperature is 
\begin{equation}
T_{\text{H}}=\frac{1}{8\pi ML_{Pl}^{2}}\exp \left( -\frac{1}{2}W\left( -%
\frac{1}{e}\left( \frac{M_{0}}{M}\right) ^{2}\right) \right) .  \label{temp}
\end{equation}%
On substituting Eq.$\left( \ref{min}\right) $ into Eq.$\left( \ref{temp}%
\right) $ we obtain the following black hole maximum temperature 
\begin{equation}
T_{H}^{max}=\frac{T_{Pl}}{2\pi \sqrt{2}\alpha }.  \label{tempmax}
\end{equation}%
The corrections to the standard Hawking temperature are obtained by
expanding Eq.$\left( \ref{temp}\right) $ in terms of $\frac{1}{e}(M_{0}/M)$.
Indeed we obtain 
\begin{equation}
T_{\text{H}}\simeq \frac{1}{8\pi ML_{Pl}^{2}}\left( 1+\frac{1}{2e}\left( 
\frac{M_{0}}{M}\right) ^{2}+\frac{5}{8e^{2}}\left( \frac{M_{0}}{M}\right)
^{4}+\frac{49}{48e^{3}}\left( \frac{M_{0}}{M}\right) ^{6}+\ldots \right) .
\label{temp-exp}
\end{equation}%
The variation of the Hawking temperature, Eq.$\left( \ref{temp}\right) $,
with the mass of the black hole is shown in figure 1.

It is interesting to inverse Eq.$\left( \ref{temp}\right) $ and write the
mass of the black hole as a function of the temperature 
\begin{equation}
M=\frac{1}{8\pi T_{H}L_{Pl}^{2}}\exp \left( \frac{1}{2}\left( \frac{T_{H}}{%
T_{H}^{max}}\right) ^{2}\right) .  \label{rt}
\end{equation}%
This relation shows that\ for temperatures larger than $T_{H}^{max}$, the
black hole mass increases with temperature. In our framework, such a
behavior is forbidden by the cut-off brought by GUP$^{\ast }$. However, in
the noncommutative approach to radiating black hole, this behavior is
allowed because of a lack of a generalized uncertainty principle \cite%
{nicolini}.

\begin{center}
\includegraphics[height=10cm,
width=10cm]{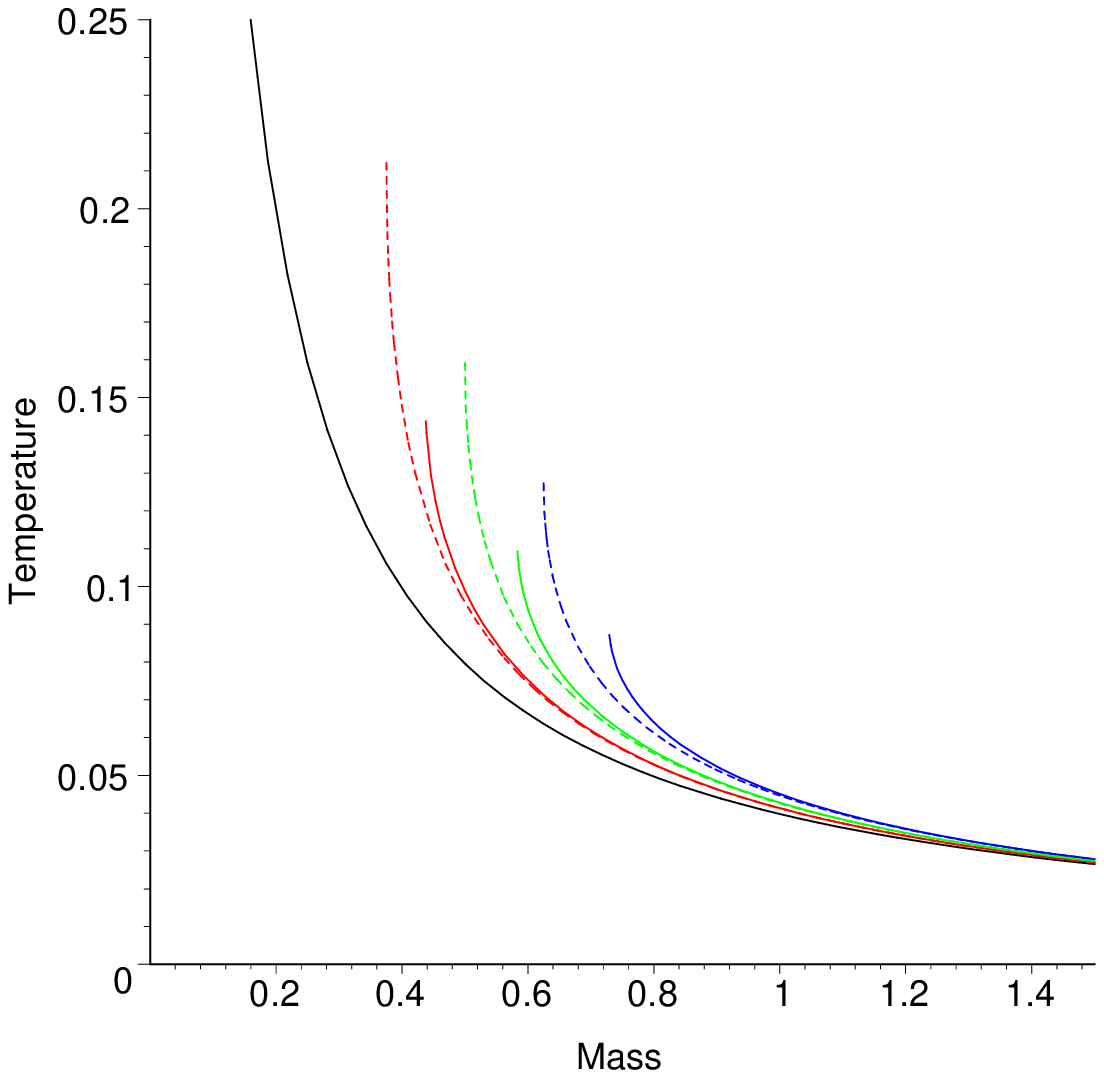}
\end{center}

Figure 1: {\small The temperature versus BH mass. \ From left to right: the
Hawking result (black solid line), GUP\ (doted line) and GUP}$^{\ast }$ 
{\small results\ (solid line) for }$\alpha ${\small =0.75 }$\left( \text{red}%
\right) ${\small , }$\alpha ${\small =1 }$\left( \text{green}\right) $%
{\small , }$\alpha ${\small =1.25 }$\left( \text{blue}\right) ${\small \
respectively.}

\bigskip 

We turn now to the calculation of the micro canonical entropy of a large
black hole. In the standard situation the entropy is proportional to the
black hole horizon-area. Following heuristic considerations due to
Bekenstein, the minimum increase of the area of a black hole absorbing a
classical particle of energy $E$ and size $R$ is given by $\left( \Delta
A\right) _{\text{0}}\simeq 4L_{Pl}^{2}\left( \ln 2\right) ER$. At the
quantum mechanical level the size and the energy of the particle are
constrained to verify $R\sim 2\delta X$ and $E\sim \delta P$. Then we have $%
\left( \Delta A\right) _{\text{0}}\simeq 8L_{Pl}^{2}\left( \ln 2\right)
\delta X\delta P.$

Extending this approach to the case with GUP$^{\ast }$ and using near
horizon geometry considerations, we obtain

\begin{equation}
\left( \Delta A\right) _{\text{0}}\approx 4L_{Pl}^{2}\ln 2\exp \left( -\frac{%
1}{2}W\left( -\frac{1}{e}\frac{A_{0}}{A}\right) \right) ,
\end{equation}%
where $A=4\pi \left( \delta X\right) ^{2}$ and $A_{\text{0}}=4\pi \left(
\delta X\right) _{\text{0}}^{2}$ are respectively the horizon area and
minimum horizon area of the black hole. With the aid of the Bekenstein
calibration factor for the minimum increase of entropy $\left( \Delta
S\right) _{\text{0}}=\ln 2$ we have 
\begin{equation}
\frac{dS}{dA}\simeq \frac{\left( \Delta S\right) _{\text{0}}}{\left( \Delta
A\right) _{\text{0}}}=\frac{1}{4L_{Pl}^{2}}\exp \left( \frac{1}{2}W\left( -%
\frac{1}{e}\frac{A_{0}}{A}\right) \right) .
\end{equation}%
Before integrating over $A$ we note that the existence of a minimum horizon
area enforces us to set the lower limit of integration as $A_{0}$. Then the
entropy, up to a irrelevant constant, is%
\begin{equation}
S\simeq \frac{1}{4L_{Pl}^{2}}\int_{A_{\text{0}}}^{A}\exp \left( \frac{1}{2}%
W\left( -\frac{1}{e}\frac{A_{0}}{A}\right) \right) dA..  \label{s1}
\end{equation}%
The relation $e^{\frac{W(x)}{2}}=\sqrt{x/W(x)}$ allows us to write Eq.$%
\left( \ref{s1}\right) $ as 
\begin{equation}
S=\frac{A_{0}}{4eL_{Pl}^{2}}\text{\textbf{PV}}\int_{-\frac{1}{e}}^{-\frac{1}{%
e}\frac{A_{0}}{A}}y^{-\frac{3}{2}}\left[ W\left( y\right) \right] ^{-\frac{1%
}{2}}dy,
\end{equation}%
where \textbf{PV }means the Cauchy principal value of the integral. Setting $%
y=-\frac{1}{e}\frac{A_{0}}{A}$ and performing the integration \ we obtain
the GUP$^{\ast }$-corrected black hole entropy 
\begin{equation}
S=\frac{A_{0}}{8eL_{Pl}^{2}}\left\{ \text{Ei}\left( -\frac{1}{2}W\left( -%
\frac{1}{e}\frac{A_{0}}{A}\right) \right) -2\left( -\frac{1}{e}\frac{A_{0}}{A%
}W\left( -\frac{1}{e}\frac{A_{0}}{A}\right) \right) ^{-\frac{1}{2}}-2\sqrt{e}%
-\text{Ei}\left( \frac{1}{2}\right) \right\} ,  \label{Entropy}
\end{equation}%
where $\text{Ei}\left( x\right) $ is the exponential function.

Expanding Eq.$\left( \ref{Entropy}\right) $ in the parameter $\frac{1}{e}%
\left( A_{0}/A\right) $ we have 
\begin{equation}
S=\left\{ \frac{A}{4L_{Pl}^{2}}-\frac{A_{0}}{8L_{Pl}^{2}e}\ln \frac{A}{A_{0}}%
+\frac{3\pi \alpha ^{2}}{16e}\left( \frac{A_{0}}{A}\right) +\frac{25\pi
\alpha ^{2}}{192e^{2}}\left( \frac{A_{0}}{A}\right) ^{2}+\frac{343\pi \alpha
^{2}}{2304e^{3}}\left( \frac{A_{0}}{A}\right) ^{3}+\ldots +\text{C}\right\} ,
\label{entropy}
\end{equation}%
where the constant is given by 
\begin{equation}
\text{C}=\frac{A_{0}}{8L_{Pl}^{2}e}\left\{ \gamma -1-2\ln \left( 2e\right) -2%
\sqrt{e}-\text{Ei}\left( \frac{1}{2}\right) \right\} \simeq -4.60\frac{%
\alpha ^{2}}{L_{Pl}^{2}}
\end{equation}%
and $\gamma $ is the Euler constant. The dependence on the Planck length is
contained in $A_{0}\sim L_{Pl}^{2}$. We observe that we have reproduced, in
our framework with GUP$^{\ast }$, the log-area correction with a negative
sign. Other approaches like string theory, loop quantum gravity and effectif
models with GUPs and/or MDRs, lead to the same sub-leading logarithmic
correction. Setting $\rho =-\frac{\pi \alpha ^{2}}{4}$ and $\beta =\frac{%
3\pi ^{2}\alpha ^{4}}{8}$ in Eqs. $\left( \ref{temp-exp}\right) $ and $%
\left( \ref{entropy}\right) $ we obtain 
\begin{eqnarray}
T_{\text{H}} &=&\frac{M_{Pl}^{2}}{8\pi M}\left[ 1-\frac{\rho }{4\pi }\left( 
\frac{M_{Pl}}{M}\right) ^{2}+\frac{\rho ^{2}+\beta /4}{16\pi ^{2}}\left( 
\frac{M_{Pl}}{M}\right) ^{4}\right] , \\
S &=&\frac{A}{4L_{Pl}^{2}}+\rho \ln \frac{A}{L_{Pl}^{2}}+\frac{\beta
L_{Pl}^{2}}{A}.
\end{eqnarray}%
These expressions are exactly the temperature and entropy obtained in loop
quantum gravity and string theory approach quantum gravity.

\begin{center}
\includegraphics[height=10cm,
width=10cm]{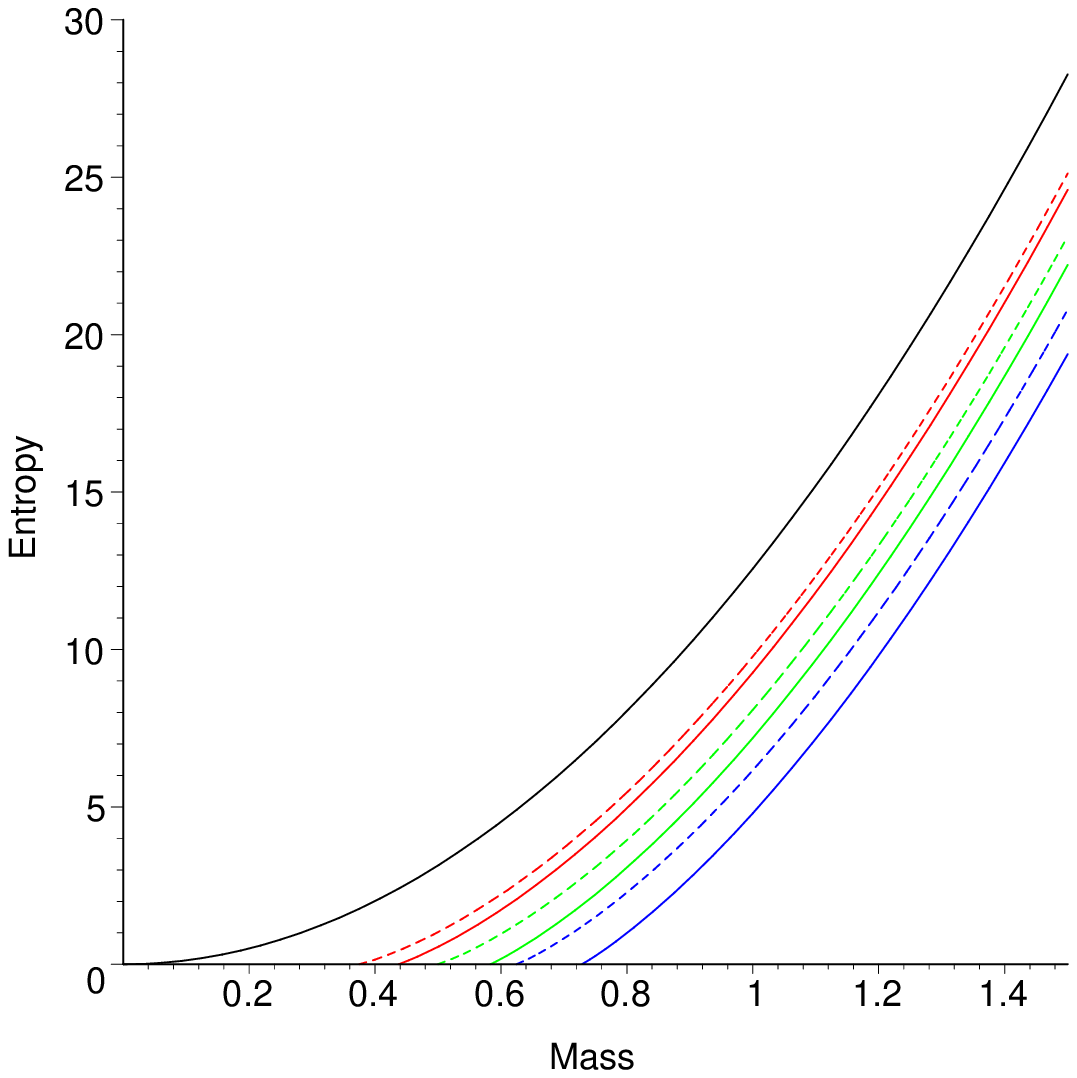}
\end{center}

Figure 2: {\small The} {\small entropy \ versus BH mass.} {\small From left
to right: the Hawking result (black solid line), GUP\ (doted line) and GUP}$%
^{\ast }${\small \ results\ (solid line) for }$\alpha ${\small =0.75 }$%
\left( \text{red}\right) ${\small , }$\alpha =1$ $\left( \text{green}\right) 
$ and $\alpha ${\small =1.25 }$\left( \text{blue}\right) ${\small \
respectively.}

From figures 1 and 2 it follows that the GUP$^{\ast }$-corrected temperature
and entropy are respectively higher and smaller than the semi classical
results. 

\section{Black holes evaporation}

As a warming to study the Hawking radiation process of the Schwarzschild
black hole, we examine the effects of GUP$^{\ast }$ on the black body
radiation spectrum. With the aid of the squeezed momentum measure given by
Eq.$\left( \ref{ferm}\right) $, which suppress the contribution of unwanted
high momenta, the energy density of a black body at temperature $T$ is
defined by 
\begin{equation}
\mathcal{E}_{\gamma }\mathcal{=}2\int d^{3}pe^{-\alpha ^{2}L_{Pl}^{2}p^{2}}%
\frac{p}{e^{\frac{p}{T}}-1}.  \label{cmp}
\end{equation}%
Using the variable $y=\beta p$ $\left( \beta =1/T\right) $ and expanding the
exponential, equation $\left( \ref{cmp}\right) $ becomes 
\begin{equation}
\mathcal{E}_{\gamma }=8\pi T^{4}\sum\limits_{n=0}^{\infty }\frac{\left(
-1\right) ^{n}}{n!}\left( \alpha T/T_{Pl}\right) ^{2n}\int_{0}^{\infty }dy%
\frac{y^{2n+3}}{e^{y}-1}.
\end{equation}%
Now with the help of the following definition of the Riemann zeta function 
\begin{equation}
\int_{0}^{\infty }dy\frac{y^{s-1}}{e^{y}-1}=\Gamma \left( s\right) \zeta
\left( s\right) ,
\end{equation}%
we obtain 
\begin{equation}
\mathcal{E}_{\gamma }=8\pi T^{4}\sum\limits_{n=0}^{\infty }\frac{\left(
-1\right) ^{n}}{n!}\left( \alpha T/T_{Pl}\right) ^{2n}\Gamma \left(
2n+4\right) \zeta \left( 2n+4\right) .  \label{energy}
\end{equation}%
This energy density is defined only for values of temperatures below some
characteristic scale. In fact Eq.$\left( \ref{energy}\right) $ is an
alternating series which converge if and only if 
\begin{equation}
\lim_{n\rightarrow \infty }\left[ \frac{1}{n!}\left( \alpha T/T_{Pl}\right)
^{2n}\Gamma \left( 2n+4\right) \zeta \left( 2n+4\right) \right] =0.
\end{equation}%
From this relation it follows that%
\begin{equation}
T<\alpha ^{-1}T_{Pl},  \label{cond}
\end{equation}%
as expected from the Gaussian damping factor in Eq.$\left( \ref{cmp}\right) $%
. However, we note that we have a stronger condition on $T$. Indeed in our
framework, the maximum temperature of the black hole is given by Eq.$\left( %
\ref{tempmax}\right) $ and it is approximately $0.1T_{Pl}$ for $\alpha $ of
order one. Then the condition on the BH temperature is rewritten as $%
T/T_{Pl}<0.1.$ For our purpose, the latter constraint allows us to cut the
series in Eq.$\left( \ref{energy}\right) $ at $n=1$. Using $\zeta \left(
4\right) =\frac{\pi ^{4}}{90}$ and $\zeta \left( 6\right) =\frac{\pi ^{6}}{%
945}$ and Eq.$\left( \ref{tempmax}\right) $ we finally obtain, from Eq.$%
\left( \ref{energy}\right) ,$ the following expression 
\begin{equation}
\mathcal{E}_{\gamma }\left( T\right) =\frac{8\pi ^{5}}{15}T^{4}\left( 1-%
\frac{15}{63}\left( \frac{T}{T_{H}^{max}}\right) ^{2}\right) .  \label{bb}
\end{equation}%
The first term is the standard Stefan-Boltzmann law while the second term is
the correction brought by GUP$^{\ast }$.

We are now ready to study the Hawking evaporation process. \ The intensity
emitted by a black hole of mass $M$ is defined by 
\begin{equation}
I=A\mathcal{E}_{\gamma }\left( T_{\text{H}}\right) ,
\end{equation}%
where $A$ is the BH horizon area. Invoking energy conservation, the
evaporation rate of the black hole is 
\begin{equation}
\frac{dM}{dt}=-A\mathcal{E}_{\gamma }\left( T_{\text{H}}\right) .
\end{equation}%
Using Eq. $\left( \ref{temp}\right) $ for the corrected Hawking temperature
we obtain%
\begin{equation}
\frac{dM}{dt}=-\frac{\gamma _{1}}{M^{2}L_{Pl}^{4}}\exp \left( -2W\left( -%
\frac{1}{e}\left( \frac{M_{0}}{M}\right) ^{2}\right) \right) \left( 1-\frac{%
8\gamma _{2}}{e\gamma _{1}}\left( \frac{M_{0}}{M}\right) ^{2}\exp \left(
-W\left( -\frac{1}{e}\left( \frac{M_{0}}{M}\right) ^{2}\right) \right)
\right) ,  \label{dm}
\end{equation}%
with $\gamma _{1}=\frac{\pi ^{2}}{480},$ $\gamma _{2}=\frac{\pi ^{2}}{16128}.
$ The deviations from the standard expression are obtained by applying a
series expansion in $\frac{1}{e}\left( M_{0}/M\right) $ 
\begin{equation}
\frac{dM}{dt}=-\frac{\gamma _{1}}{M^{2}L_{Pl}^{4}}\left( 1+\frac{2}{e}\left( 
\frac{M_{0}}{M}\right) ^{2}+\frac{4}{e^{2}}\left( 1-\frac{2\gamma _{2}}{%
e\gamma _{1}}\right) \left( \frac{M_{0}}{M}\right) ^{4}+\frac{25}{3e^{3}}%
\left( 1-\frac{72\gamma _{2}}{25e\gamma _{1}}\right) \left( \frac{M_{0}}{M}%
\right) ^{6}+\ldots \right) .  \label{DM}
\end{equation}%
The variation of the evaporation rate with the black hole mass is shown in
Figure 3. We clearly observe that the divergence for $M\rightarrow 0$ in the
standard description of the black hole evaporation process is now completely
regularized by the GUP$^{\ast }$. This regularization is also reflected by
the constraint $\left( \ref{cond}\right) $, which suppress the evaporation
process beyond the Planck temperature. This phenomenon is similar to the
prevention, by the standard uncertainty principle, of the hydrogen atom from
total collapse. In our picture, the regularization can be considered as a
dynamical effect and \ not as a consequence of any quantum symmetry in the
theory.

\begin{center}
\includegraphics[height=10cm,
width=10cm]{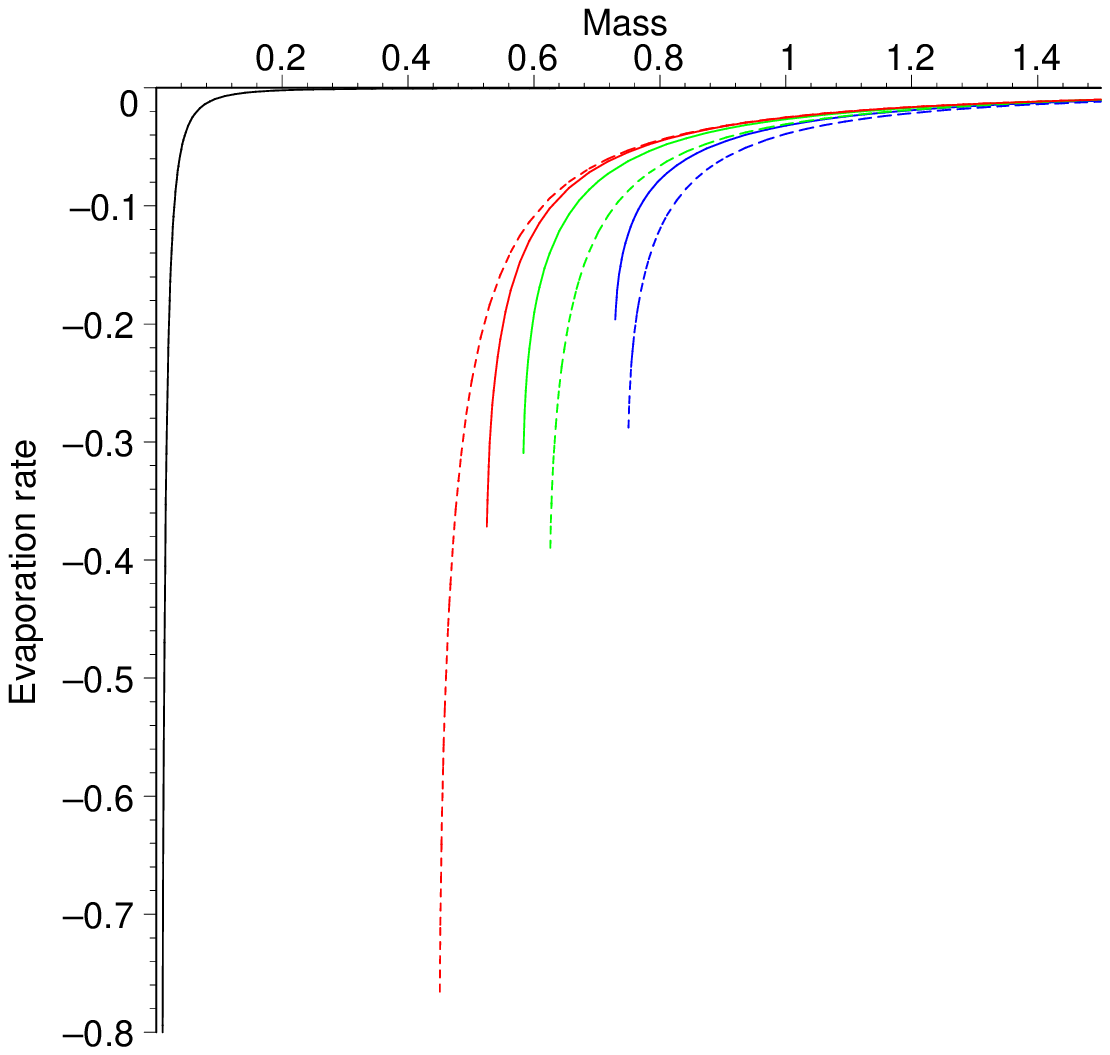}
\end{center}

Figure 3: {\small The} {\small evaporation rate versus BH mass. From left to
right: the Hawking result }$\left( \text{{\small black solid line}}\right) ,$%
{\small \ GUP (doted line) and GUP}$^{\ast }${\small \ results (solid line)
for }$\alpha =0.90${\small \ }$\left( \text{{\small red}}\right) ${\small \ }%
$,${\small \ }$\alpha =1.${\small \ }$\left( \text{{\small green}}\right) $%
{\small \ and }$\alpha =1.25${\small \ }$\left( \text{{\small blue}}\right) $%
{\small .}

On the other hand, we observe that the evaporation phase ends when the BH
mass becomes equal to $M_{0}$ with a minimum rate given by 
\begin{equation}
\left( \frac{dM}{dt}\right) _{min}=-\frac{e^{2}}{M_{0}^{2}}\left( \gamma
_{1}-8\gamma _{2}\right) M_{Pl}^{4}.  \label{rate-min}
\end{equation}%
Thus the evaporation process of a black hole with initial mass $M>M_{\text{0}%
}$ continue until the horizon radius becomes $\left( \delta X\right) _{\text{%
0}},$ leaving a massive relic referred to, in the literature, as a black
hole remnant (BHR). To find the nature of the BHR we calculate the heat
capacity defined by 
\begin{equation}
C=\frac{dM}{dT_{\text{H}}}.
\end{equation}%
Using the expression of temperature given by $\left( \ref{temp}\right) $ we
easily obtain 
\begin{equation}
C=-8\pi M^{2}L_{Pl}^{2}\left( 1+W\left( -\frac{1}{e}\left( \frac{M_{0}}{M}%
\right) ^{2}\right) \right) \exp \left( \frac{1}{2}W\left( -\frac{1}{e}%
\left( \frac{M_{0}}{M}\right) ^{2}\right) \right) .  \label{heat}
\end{equation}%
This expression vanishes when $1+W\left( -\frac{1}{e}\left( M_{0}/M\right)
^{2}\right) =0,$ whose solution is $M=M_{0}.$ We conclude that the heat
capacity of the black hole vanishes at the end point of the evaporation
process characterized by a BHR with mass  $M_{0}.$Besides the gravitational
interaction with the surrounding, the vanishing of the heat capacity reveals
the inert character of the BHRs and thus make them as potential candidates
to explain the origin of dark matter \cite{Pisin,P-adler}. Finally we note
that, as it is the case with the form of the GUP to leading order in the
Planck length, the BHRs are also a consequence of GUP$^{\ast }$ \cite%
{adler,marco}.

We have drawn the variation of the heat capacity with BH mass in figure 4.
In it we see that, the heat capacity vanishes for $M_{0}\simeq 0.50,$ $0.75$
in the case with GUP and $M_{\text{0}}\simeq 0.58,$ $0.87$ in the case with
GUP$^{\ast }.$ 

\begin{center}
\includegraphics[height=10cm,
width=10cm]{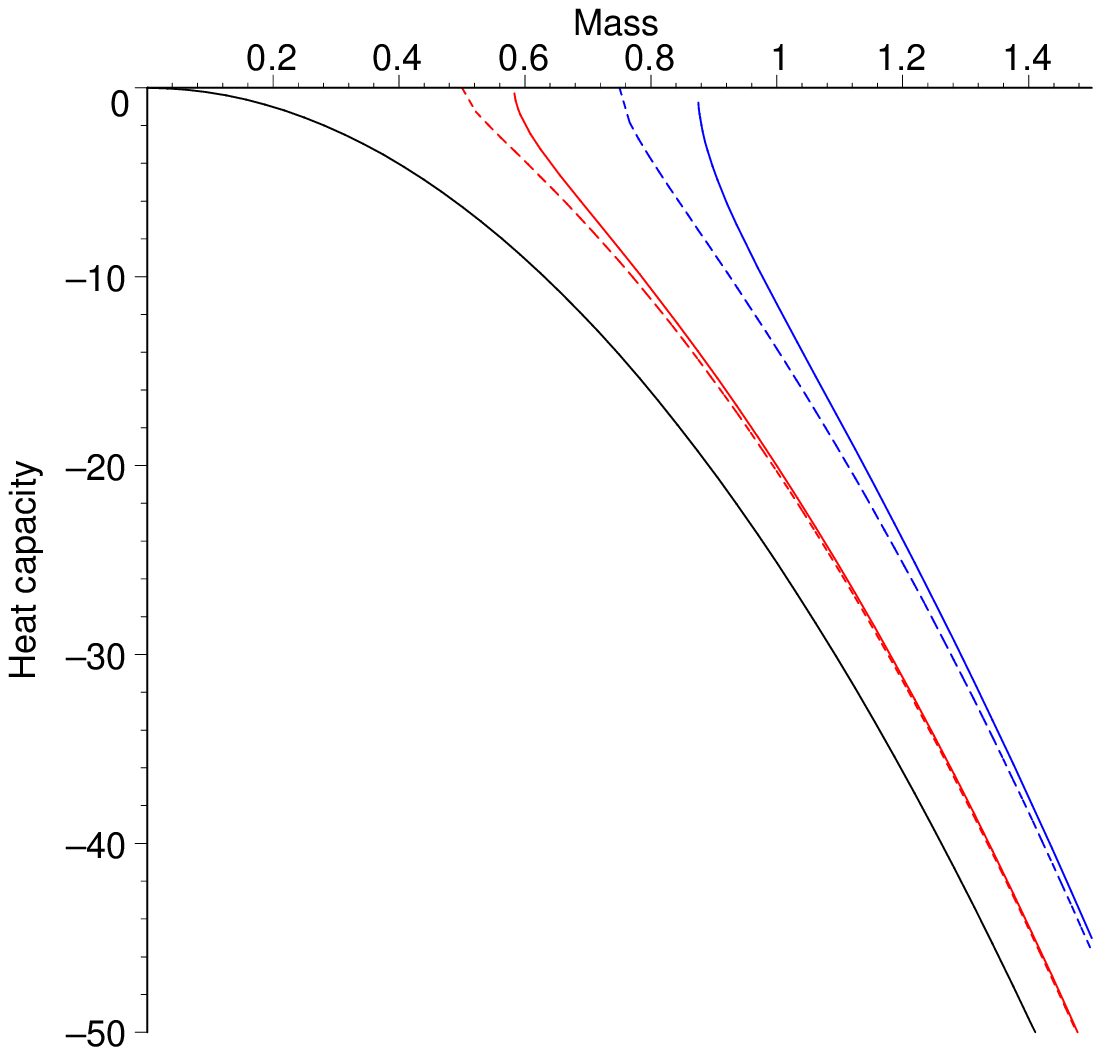}
\end{center}

Figure 4: {\small The heat capacity \ versus BH mass. From left to right:
the standard result (black solid line), GUP (doted line) and GUP}$^{\ast }$%
{\small \ results (solid line) for }$\alpha =1${\small \ }$\left( \text{%
{\small red}}\right) ${\small \ and }$\alpha =1.5${\small \ }$\left( \text{%
{\small blue}}\right) $ {\small respectively}$.$ \bigskip 

Taylor expanding Eq.$\left( \ref{heat}\right) $ we have 
\begin{equation}
C= -8\pi M^{2}L_{Pl}^{2}\left( 1-\frac{3}{2e}\left( \frac{M_{0}}{M}\right)
^{2}-\frac{7}{8e^{2}}\left( \frac{M_{0}}{M}\right) ^{4}-\frac{55}{48e^{3}}%
\left( \frac{M_{0}}{M}\right) ^{6}-\ldots \right) .  \label{c}
\end{equation}%
The standard expression of the heat capacity $C= -8\pi M^{2}$ is reproduced
in the limit of black holes with mass larger than the minimum mass $M_{0}$.
The correction terms to the heat capacity due to GUP$^{\ast }$ are all
positive indicating that the evaporation process is accelerated and leading
to a corrected decay time smaller than the decay time in the standard case.

Let us consider a black hole starting the evaporation process with a mass $M$
and ending the process with the minimum mass $M_{0}$. Using $\left( \ref{dm}%
\right) $ and the variable $y=-\frac{1}{e}\left( M_{0}/M\right) ^{2},$ the
decay time is given by

\begin{eqnarray}
t &=&(-1)^{7/2}\frac{M_{0}^{3}L_{Pl}^{4}}{2\gamma _{1}e^{3/2}}\int_{-\frac{1%
}{e}\left( M_{0}/M\right) ^{2}}^{-\frac{1}{e}}W^{-5/2}\left( y\right) e^{-%
\frac{1}{2}W\left( y\right) }dy  \notag \\
&+&(-1)^{5/2}\frac{\gamma _{2}}{2\gamma _{1}^{2}e^{1/2}}M_{0}\alpha
^{2}L_{Pl}^{2}\int_{-\frac{1}{e}\left( M_{0}/M\right) ^{2}}^{-\frac{1}{e}%
}W^{-3/2}\left( y\right) e^{-\frac{1}{2}W\left( y\right) }dy.
\end{eqnarray}%
Performing the integration we obtain 
\begin{equation}
t=\frac{M_{0}^{3}L_{Pl}^{4}}{3\gamma _{1}e^{3/2}}\left[ 4\left( 1-\epsilon
\right) \frac{\sqrt{-y}}{W\left( y\right) }-\sqrt{8\pi }\left( 1-\frac{%
\epsilon }{2}\right) \text{erf}\left( \sqrt{-\frac{W\left( y\right) }{2}}%
\right) +\frac{\sqrt{-y}}{W^{2}\left( y\right) }\right] +C,  \label{decay}
\end{equation}%
where the constant $C$ \ is the value for $y=-1/e$ and $\epsilon =\frac{%
3\gamma _{2}}{\gamma _{1}}e\alpha ^{2}\left( \frac{M_{Pl}}{M_{0}}\right)
^{2}\sim 10^{-6}$ for $\alpha $ of order one. Ignoring $\epsilon $ and
performing a series expansion in $y$ \ we have 
\begin{equation}
t=\frac{M^{3}L_{Pl}^{4}}{3\gamma _{1}}\left( 1-\frac{6}{e}\left( \frac{M_{0}%
}{M}\right) ^{2}-\frac{2}{e^{2}}\left( \frac{M_{0}}{M}\right) ^{4}+\frac{1}{%
3e^{3}}\left( \frac{M_{0}}{M}\right) ^{6}+\ldots \right) 
\end{equation}%
Then to first order in $\frac{1}{e}\left( M_{0}/M\right) $ the relative
correction to the decay time is 
\begin{equation}
\frac{\Delta t}{t_{0}}=-\frac{6}{e}\left( \frac{M_{0}}{M}\right) ^{2},
\label{time}
\end{equation}%
where $t_{0}=\frac{M^{3}L_{Pl}^{4}}{3\gamma _{1}}$ is the decay time without
GUP$^{\ast }$. From Eq.$\left( \ref{time}\right) $ , it follows that black
holes with GUP$^{\ast }$ are hotter and decay  faster than in the standard
case.\newline

Let us now turn to a comparison of the corrected BH thermodynamics with GUP$%
^{\ast }$ with the corrections brought by the GUP to leading order in the
Planck length. Since our comparison is quantitative we use the Planck units.
Repeating the same calculations as above with the GUP given by Eq.$\left( %
\ref{logup}\right) $, the temperature, the entropy and the heat capacity of
the black hole are respectively given by

\begin{equation}
T_{GUP}=\frac{M}{\pi \alpha ^{2}}\left( 1-\sqrt{1-\frac{\alpha ^{2}}{4M^{2}}}%
\right) ,  \label{Tgup}
\end{equation}%
\begin{equation}
S_{GUP}=2\pi M^{2}\left( 1+\sqrt{1-\frac{\alpha ^{2}}{4M^{2}}}\right) -\frac{%
\pi \alpha ^{2}}{8}\ln \left( \frac{8M^{2}}{\alpha ^{2}}\left( 1+\sqrt{1-%
\frac{\alpha ^{2}}{4M^{2}}}\right) -1\right) -\frac{\pi \alpha ^{2}}{8}
\label{Sgup}
\end{equation}%
and

\begin{equation}
C_{GUP}=\pi \alpha ^{2}\frac{\sqrt{1-\frac{\alpha ^{2}}{4M^{2}}}}{\sqrt{1-%
\frac{\alpha ^{2}}{4M^{2}}}-1}.  \label{Cgup}
\end{equation}%
The minimum black hole mass and maximum temperature allowed by GUP are $%
M_{0}=\left( \delta X\right) _{0}/2=\frac{\alpha }{2}$ and $T_{\text{max}%
}=1/2\pi \alpha .$ In figures 1, 2, 3 and 4 we have plotted, besides the
results obtained with GUP$^{\ast }$, the variation of temperature, entropy,
evaporation rate and heat capacity with GUP as functions of the black hole
mass for different values of the parameter $\alpha .$ Figure 2 shows, that
in the scenario with GUP$^{\ast },$ the BH entropy \ decreases compared to
the entropy in the standard case and the scenario with GUP. \ This reveals
the deeper quantum nature of the black hole in the scenario with GUP$^{\ast
} $. Thus quantum effects become manifest at an earlier stage of the
evaporation phase than was predicted by the semi classical Hawking analysis 
\cite{cavag} and the GUP analysis $\cite{marco}.$

The calculation of the evaporation rate in the framework with GUP requires a
careful analysis. In all the calculations done until now, the validity of
the Stefan-Boltzmann law is assumed, ignoring the UV cut-off implemented by
GUP. However, it was pointed in \cite{c9} that the effect of the GUP should
be also reflected in a modification of the de Broglie wave length relation

\begin{equation}
\lambda =\frac{1}{2p}\left( 1+\alpha ^{2}p^{2}\right) .
\end{equation}%
This relation must be translated into a modification of the momentum measure
such that the contributions of high momenta are suppressed. As shown in \cite%
{c222}, the GUP to leading order in the Planck length leads to a squeezing
of the momentum measure by a factor $\frac{1}{\left( 1+\alpha
^{2}L_{Pl}^{2}p^{2}\right) }.$ Then following the same calculation leading
to Eq.$\left( \ref{energy}\right) ,$ the energy density of a black body with
GUP is

\begin{equation}
\mathcal{E}_{\gamma }\mathcal{=}2\int \frac{dp^{3}}{\left( 1+\alpha^{2}
L_{Pl}^{2}p^{2}\right) }\frac{p}{e^{\frac{p}{T}}-1}.
\end{equation}%
Performing the integral and using the same argument as before, we obtain the
expression given by Eq. $\left( \ref{bb}\right) .$ We note, that in a recent
calculation of the Stefan-Boltzmann law with GUP \cite{c14},\ the sign of
the correction term is positive, in contradiction with the role of the UV
cut-off implemented by the GUP.

The correct evaporation rate with GUP is then given by

\begin{equation}
\left( \frac{dM}{dt}\right) _{GUP}=-\frac{128\pi ^{2}M^{6}}{15\alpha ^{8}}%
\left( 1-\sqrt{1-\frac{\alpha ^{2}}{4M^{2}}}\right) ^{4}+\frac{1024\pi
^{2}M^{8}}{63\alpha ^{10}}\left( 1-\sqrt{1-\frac{\alpha ^{2}}{4M^{2}}}%
\right) ^{6}.
\end{equation}%
In figure 3 we observe that the evaporation process with GUP is retarded
compared to the process with GUP$^{\ast }$ and that the process ends at a
mass $M_{\text{0}}=\alpha /2$ with a minimum rate given by

\begin{equation}
\left( \frac{dM}{dt}\right) _{min,GUP}=-\frac{32\pi ^{2}}{1260 M_{0}^{2}},
\end{equation}%
which is greater (in absolute value) than the one obtained with GUP$^{\ast }$%
.

In table 1 we show the GUP and GUP$^{\ast }$-corrected thermodynamics of two
black holes with initial mass equal to $2M_{Pl}$ and $5M_{Pl}$ for $\alpha
=1.$ The first row gives the semi classical Hawking results. The second row
gives the GUP-corrected results and the third row the GUP$^{\ast }-$%
corrected ones. It is interesting to note that, in the scenario with GUP$%
^{\ast },$ the final stage of the evaporation phase is a remnant with a mass
larger than the one obtained with GUP and that the decay time is drastically
reduced. In a scenario with extra dimensions, these results may have
important consequences on possible black holes production at particle
colliders and in ultrahigh energy cosmic ray (UHECR) air-showers.

Finally, let us notice that the corrections to the black hole thermodynamics
become indistinguishable in the two version of GUP in the limit of large
mass and small values of $\alpha .$ However, for growing values of the
parameter $\alpha ,$ corresponding to strong gravitational effects, the
predictions of the two GUPs concerning the entropy become different even for
massive black holes.

\bigskip

Table 1. {\small GUP and GUP$^{\ast }$-corrected thermodynamics for two BHs
with mass $M=2$ and $M=5$ (in Planck units). The deviations from the Hawking
results are also given.}

\begin{center}
$M=2$

\begin{tabular}{|c|c|c|c|c|c|}
\hline
${\small \alpha }$ & {\small Minimum mass} & {\small Initial temp} & {\small %
Final temp} & {\small Decay time} & {\small Entropy} \\ \hline
{\small 0} & {\small -} & {\small 0.019} & ${\small \infty }$ & {\small %
129.69} & {\small 50.27} \\ \hline
{\small 1.0} & {\small 0.5} & {\small 0.020} \ $\left( +3\%\right) $ & 
{\small 0.16} & {\small 111.92} $\left( -14\%\right) $ & {\small 44.66} $%
\left( -11\%\right) $ \\ \hline
{\small 1.0} & {\small 0.58} $\left( +16\%\right) $ & {\small 0.020} $\left(
+3\%\right) $ & {\small 0.11} $\left( -31\%\right) $ & {\small 3.33} $\left(
-97\%\right) $ & {\small 43.73} $\left( -13\%\right) $ \\ \hline
\end{tabular}
\end{center}

\bigskip

\begin{center}
$M=5$

\begin{tabular}{|c|c|c|c|c|c|}
\hline
${\small \alpha }$ & {\small Minimum mass} & {\small Initial temp} & {\small %
Final temp} & {\small Decay time} & {\small Entropy} \\ \hline
{\small 0} & {\small -} & {\small 0.008} & ${\small \infty }$ & {\small %
2026.42} & {\small 314.16} \\ \hline
{\small 1.0} & {\small 0.5} & {\small 0.008} & {\small 0.16} & {\small %
1976.60} $\left( -2.5\%\right) $ & {\small 307.10} $\left( -2\%\right) $ \\ 
\hline
{\small 1.0} & {\small 0.58} $\left( +16\%\right) $ & {\small 0.008} & 
{\small 0.11} $\left( -31\%\right) $ & {\small 22.17} $\left( -99\%\right) $
& {\small 306.18} $\left( -2.2\%\right) $ \\ \hline
\end{tabular}
\end{center}

\section{Conclusion}

In this paper we have studied how black holes thermodynamic parameters are
affected by a GUP to all orders in the Planck length. We have obtained exact
analytic expressions for the Hawking temperature and entropy. Particularly
we found that a black hole with a mass smaller than a minimum mass do not
exist. The existence of a energy scale which is one order below the Planck
scale allowed us to calculate, to leading order, the deviations from the
standard Stefan-Boltzmann law. Then we investigated the Hawking radiation
process of the Schwarzschild black hole and shown that at the end of the
evaporation phase a inert massive relic continue to exist as a black hole
remnant (BHR) with zero entropy, zero heat capacity and non zero finite
temperature. For completeness, we have also compared our results with the
semi classical results and the predictions of the GUP to leading order in
the Planck length. In particular, we have shown that the entropy in our
framework is smaller than the entropy in the standard case and with GUP. We
have also made the correct calculation of the evaporation rate with GUP. \
Finally, we have shown that black holes with the form of GUP used in this
paper are hotter, shorter-lived and tend to evaporate less than black holes
in the semi classical and the GUP to leading order pictures. On the other
hand, we have found that the predictions of the GUP to all orders in the
Planck length and the GUP to leading order in the Planck length, concerning
the entropy, become different for strong gravitational effects and large
black holes mass, suggesting a further investigation of the early universe
thermodynamics in the framework with the GUP to all orders in the Planck
length. In a future work we will examine the effects of the GUP to all
orders in the Planck length on black holes thermodynamics in a scenario with
extra dimensions. \newline
\newline

\textbf{Acknowledgments}

The author thanks the referees for their comments and valuable remarks.

\end{document}